\documentclass[aps,prb,showpacs,tightenlines, twocolumn,secnumarabic,nofootinbib,nobibnotes,superscriptaddress]{revtex4-1}
\usepackage{amsmath,amssymb,amsfonts,bm}
\usepackage{graphicx}
\usepackage{epstopdf}
\usepackage{dcolumn}
\usepackage{mathrsfs}  
\usepackage[colorlinks=true,linkcolor=blue,citecolor=blue, urlcolor=blue,bookmarks=false]{hyperref}

\begin{document}
\title{Single magnetic impurity in tilted Dirac surface states }
 \author{Jin-Hua Sun}
 \email{sunjinhua@nbu.edu.cn}
 \author{Lu-Ji Wang}
 \author{Xing-Tai Hu}
 \affiliation{
 	 Department of Physics, Ningbo University, Ningbo, China}
 \author{Lin Li}
  \affiliation{
 	 Department of Physics, Southern University of Science and Technology, Shenzhen, China}
 \author{Dong-Hui Xu}
 \email{donghuixu@hubu.edu.cn}
 \affiliation{
 	 Department of Physics, Hubei University, Wuhan, China}
  
\begin{abstract}
	We utilize variational method to investigate the Kondo screening of a spin-1/2 magnetic impurity in tilted Dirac surface states with the Dirac cone tilted along the $k_y$-axis. We mainly study about the effect of the tilting term on the binding energy and the spin-spin correlation between magnetic impurity and conduction electrons. 
	The binding energy has a critical value while the Dirac cone is slightly tilted. However, as the tilting term increases, the density of states near the Dirac node becomes significant, such that the impurity and the host material always favor a bound state. The diagonal and the off-diagonal terms of the spin-spin correlation between the magnetic impurity and conduction electrons are also studied. Due to the spin-orbit coupling and the tilting of the spectra, various components of spin-spin correlation show very strong anisotropy in coordinate space, and are of power-law decay with respect to the spatial displacements.
\end{abstract}

\maketitle

\section{Introduction}

Topological semimetals\cite{armitage2017} host Dirac or Weyl fermions in the bulk and have attracted lots of theoretical and experimental research interests in recent years. The Dirac or Weyl fermions found in condensed matter physics are quasiparticles which do not have to obey the Lorentz invariance, indicating that the band structure in the momentum space can be anisotropic. Type-II Dirac or Weyl fermions\cite{soluyanov2015, Xu2015prl} are obtained when Dirac or Weyl cones are tilted strongly in the momentum space, that the electron and hole pockets co-exist with the Dirac or Weyl nodes. Type-II Weyl fermions are predicted to exist in many materials, such as  $\text{WTe}_2$\cite{soluyanov2015}, $\text{MoTe}_2$\cite{Sunyan2015,Wang2016}, $\text{Ta}_3\text{S}_2$ and $\text{LaAlGe}$\cite{XuSY2017}. More recently, it has been reported that $\text{PdTe}_2$\cite{Noh2017, Fei2017} and $\text{PtTe}_2$\cite{yan2017} are type-II Dirac semimetals which host tilted Dirac cones in three-dimensions.

Except for the type-II topological semimetals mentioned above, one can also obtain tilted Dirac or Weyl cones in two-dimensions \cite{Chiu2017, Nagaosa2016}. The tilted anisotropic Dirac cones have been found in the $8-pmmn$ borophene\cite{Lopez2016} and the organic semiconductor $\alpha$-$\text{(BEDT-TTF)}_2\text{I}_3$\cite{Goerbig2008,Hirata2017}. In particular, it has been proposed that the crystal symmetries can give rise to type-II Dirac surface states\cite{Chiu2017} which are characterized by tilted Dirac cones with helical spin polarization and open electron and hole pockets touching at the Dirac point.

The purpose of this paper is to investigate the properties of the Kondo screening in two-dimensional (2D) tilted Dirac surface states with helical spin polarization. The Kondo problem is an important issue in condensed matter physics and has been widely studied by using various methods \cite{Krishna1980,tsvelick1984,Andrei1984,Zhang1983,Coleman1984,read1983,Kuramoto1983,Gunnarsson1983,affleck1990}. The Kondo problem as well as the RKKY interactions in systems with isotropic Dirac cones have been studied intensively since the discoveries of graphene and topological insulators\cite{Chang2015,Ulloa2016,Shun2016,Zheng2016}. At half-filling, the density of states (DOS) of the Dirac fermions vanishes, and the problem of a magnetic impurity in such systems falls into the category of pseudo-gap Kondo problem \cite{Gonzalez1998,Fritz2004,Vojta2004}. There exists a critical value of hybridization for the impurity and the conduction electrons to form a bound state\cite{Feng2010, shirakawa2014}. For tilted Dirac surface states, due to the co-existance of spin-orbit coupling and the anisotropy of band structure, the spin-spin correlations in both the spin and coordinate spaces show rich features and are much more interesting than those in normal metals.

In this paper, we systematically study the binding energy and real space spin-spin correlations of a magnetic impurity in titled Dirac surface states. We use the variational method, and compare the results with those obtained in conventional 2D helical metals. The variational method we apply has been used to study the ground state of the Kondo problem in normal metals \cite{Gunnarsson1983,Varma1976}, antiferromagnet \cite{Aji2008}, 2D helical metals \cite{Feng2010}, 3D Weyl semimetals\cite{Jinhua2015}, and the Fermi arc surface states of Weyl semimetals\cite{Ma2017}.

The paper is organized as follows. We present the model and dispersion relation in Sec. \ref{Sec:Hamiltonian}. 
In Sec. \ref{Sec:selfconsist}, we apply the variational method to study the binding energy. In Sec. \ref{Sec:sscorr}, we investigate the spin-spin correlation between the magnetic impurity and the conduction electrons in tilted Dirac surface states. Two cases are mainly studied: (1) $v_x=v_y$, $v_t\neq 0$ and (2) $v_x\neq v_y$, $v_t \neq 0$, where $v_x$, $v_y$ are the velocities along the $k_x$- and $k_y$-axis and $v_t$ is the tilting term. The results are compared with the counterparts in a two dimensional helical metal ($v_x=v_y$, $v_t=0$).
Finally, the discussions and conclusions are given in
Sec. \ref{conclusion}.

\section{Hamiltonian}\label{Sec:Hamiltonian}

We use the Anderson impurity model to study the Kondo screening of a spin-1/2 magnetic impurity in tilted Dirac surface states. The model Hamiltonian contains three parts: the kinetic energy term $H_0$ of the tilted Dirac cone, the impurity Hamiltonian $H_d$, and the hybridization between the magnetic impurity and the tilted Dirac surface states $H_V$. The Hamiltonian reads

\begin{equation}\label{Eq:tilt_Dirac}
\begin{aligned}
H=H_0 + H_d + H_V.
\end{aligned}
\end{equation}

The Hamiltonian of a tilted Dirac cone in a 2D plane is given by  \cite{Zabolotskiy2016,SK2017}
\begin{equation}\label{Eq:tilt_Dirac}
\begin{aligned}
H_0 = \sum_{\mathbf{k}} h_0(\mathbf{k})=\sum_\mathbf{k} \Psi_{\mathbf{k}}^\dagger \left(v_x k_x \sigma_x + v_y k_y \sigma_y + v_t k_y \sigma_0\right)\Psi_{\mathbf{k}},
\end{aligned}
\end{equation}
where $\sigma_x$, $\sigma_y$ are the spin Pauli matrices and $\sigma_0$ is the identity matrix. $\Psi_{\mathbf{k}}\equiv \{c_{\mathbf{k}\uparrow}, c_{\mathbf{k}\downarrow}\}^T$ and $\Psi_{\mathbf{k}}^\dagger = \{c_{\mathbf{k}\uparrow}^\dagger, c_{\mathbf{k}\downarrow}^\dagger\}$, where $c_{\mathbf{k}\sigma}^\dagger$ ($c_{\mathbf{k}\sigma}$) creates (annihilates) an spin-$\sigma$ electron with momentum $\mathbf{k}$. $v_x$ and $v_y$ are the velocity along the $k_x$ and $k_y$ axes, respectively.
When $v_t = 0$ and $v_x = v_y$, the dispersion relation is exactly the same as a single Dirac cone in graphene or in a 2D helical metal. The non-zero $v_t$ tilts the Dirac cone, and if $v_x \neq v_y$ extra anisotropy is induced in the system, such that the real space spin-spin correlation between a magnetic impurity and the conduction electrons shall be affected accordingly.   \\

\begin{figure}[htpb]
	\begin{center}
		\includegraphics[width=8cm]{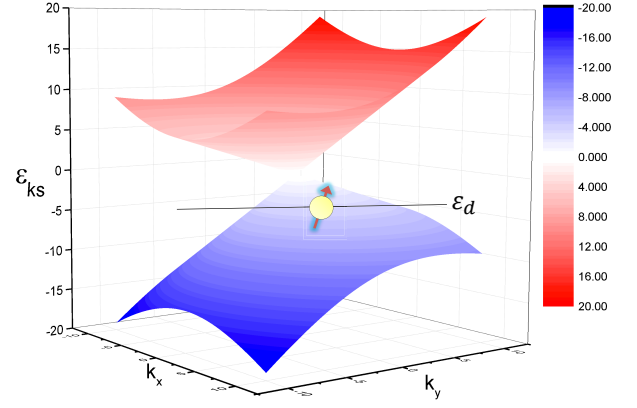}
	\end{center}
	\caption{(Color online). The band structure of tilted Dirac cone for $v_x = v_y = 1.0$ and $v_t = 0.5$. $\epsilon_d$ is the impurity energy level which is below the Fermi surface. The dispersion relation is tilted along the $k_y$-axis due to the non-zero $v_t$ term.  } \label{Fig:0_dispersion.pdf}
\end{figure}

The single particle eigenenergy writes
\begin{equation}\label{Eq:tilt_dispersion}
\begin{aligned}
\epsilon_{ks} = k_y v_t - s \sqrt{k_x^2v_x^2 + k_y^2v_y^2},
\end{aligned}
\end{equation}
where $s=\{+, -\}$ refer to the valence and the conduction bands. The dispersion relation for $v_x=v_y=1.0$ and $v_t=0.5$ is shown in Fig. \ref{Fig:0_dispersion.pdf}. If $v_t=0$, the spectrum is isotropic in the 2D plane. The non-zero $v_t$ tilts the Dirac cone along the $k_y$-axis. We can see that the DOS is still zero for a small $v_t$, but as $v_t$ increases, the DOS at half-filling will become finite. In this present paper, we may study about the case with relatively small $v_t$, such that the DOS at half-filling is still zero while the spectra become anisotropic due to the tilting term.

The eigenstates are given by $\{\{ -e^{-i\theta_\mathbf{k}}, 1\}, \{e^{-i\theta_\mathbf{k}},1\}\}$, where $\theta_\mathbf{k} \equiv \text{arctan}(-k_yv_y/k_xv_x)$. Then one can define a unitary matrix to diagonalize $h_0(\mathbf{k})$ as
\begin{equation}
\begin{aligned}
U =\frac{1}{\sqrt{2}}\left(
\begin{array}{cc}
e^{-\frac{i\theta_\mathbf{k}}{2}} & -e^{\frac{i\theta_\mathbf{k}}{2}}  \\
e^{-\frac{i\theta_\mathbf{k}}{2}} &  e^{ \frac{i\theta_\mathbf{k}}{2}}
\end{array}
\right) .\\
\end{aligned}
\end{equation}
The eigenstates of the tilted Dirac cone is given by
\begin{equation}
\begin{aligned}
\{\gamma_{\mathbf{k}+}, \gamma_{\mathbf{k}-}\}^T = U\{c_{\mathbf{k}\uparrow}, c_{\mathbf{k}\downarrow}\}^T,\\
\end{aligned}
\end{equation}
and then $H_0$ in its diagonal basis writes
\begin{equation}
\begin{aligned}
H_0 = \sum_\mathbf{k} h_0(\mathbf{k}) =  \sum_{\mathbf{k}s} \epsilon_{\mathbf{k}s}\gamma_{\mathbf{k}s}^\dagger \gamma_{\mathbf{k}s}, \ \  (s=\{+, -\}). \\
\end{aligned}
\end{equation}

The local impurity Hamiltonian is given by

\begin{equation}\label{Eq:Hamil_Impu}
\begin{aligned}
H_{d} &=(\epsilon_d -\mu) \sum _{\sigma }d_{\sigma }{}^\dagger d_{\sigma }+\text{U}d_{\uparrow }{}^\dagger d_{\uparrow }d_{\downarrow
}{}^\dagger d_{\downarrow },\\
\end{aligned}
\end{equation}
$d_{\uparrow(\downarrow)}^\dagger$ and $d_{\uparrow(\downarrow)}$ are the creation and annihilation operators of the
spin-up (spin-down) state on the impurity site.
$\epsilon_d$ is the impurity energy level, $U$ is the on-site Coulomb repulsion. We may assume that $\epsilon_d$ is slightly below the chemical potential and $U$ is finite but very large, such that  $\epsilon_d<\mu\ll \epsilon_d+U$, that the impurity is always singly occupied
with a local moment, and the impurity energy shall be $\epsilon_d-\mu$.

The hybridization between the electrons on the magnetic impurity site and in the tilted Dirac cone is described by
\begin{equation}\label{Eq:Hv}
\begin{aligned}
 H_{V}&=\sum _{\mathbf{k}\sigma}V_\mathbf{k}\left(c_{\mathbf{k}\sigma}^\dagger d_{\sigma} + d_{\sigma}^\dagger c_{\mathbf{k}\sigma}  \right) =\sum_{\mathbf{k}s} V_{\mathbf{k}}(\gamma_{\mathbf{k}s}^\dagger d_{\mathbf{k}s} + \gamma_{\mathbf{k}s}d_{\mathbf{k}s}^\dagger),
\end{aligned}
\end{equation}
$V_k$ is the hybridization strength, and we assume that the electrons on the magnetic impurity is equally coupled to the conduction and valence bands. The momentum space impurity operators $d_{\mathbf{k}s}$ are connected to the original ones $d_{\sigma}$ through the following unitary transformation
\begin{equation}
\begin{aligned}
\{d_{\mathbf{k}+}, d_{\mathbf{k}-}\}^T = U\{d_\uparrow, d_{\downarrow}\}^T.\\
\end{aligned}
\end{equation}

We assume that the hybridization only occurs between the magnetic impurity and the conduction electrons on the same location in coordinate space. Hence in the following, the hybridization strength $V_{\mathbf{k}}$ is in fact momentum-independent.

\section{The self-consistent calculation }\label{Sec:selfconsist}

First we may assume $H_V=0$, which is the simplest case that the magnetic impurity and the host material is completely decoupled from each other. The ground state of $H_0$ is given by
\begin{equation}
\begin{aligned}
|\Psi_0 \rangle =\prod _{\mathbf{k}s} \gamma _{\mathbf{k}s}^{\dagger}|0\rangle , 
\end{aligned}
\end{equation}
where the product runs over all the states below the Fermi surface, and $s=\{+,-\}$ refer to the valence and the conduction bands in the tilted Dirac cone.
If we consider about singly occupied impurity, and ignore the hybridization between conduction electrons and the magnetic impurity, the total energy of the system is just the sum of the bare impurity energy and the total energy of the tilted Dirac cone,
\begin{equation}
\begin{aligned}
E_0=\epsilon _d -\mu +\sum _{\mathbf{k}s}(\epsilon_{\mathbf{k}s}-\mu).
\end{aligned}
\end{equation}

In order to investigate the eigenstate property, we utilize a trial wavefunction approach. The Coulomb repulsion $U$ is assumed to be a finite but very large value, and $\epsilon_d$ is below the chemical potential, such that the impurity site is always singly occupied.
If the hybridization interaction is taken into account, the band electron states and the localized states are combined. 
According to the most right side of Eq. \ref{Eq:Hv}, the hybridization term only involves the band states and the impurity states with the same indices $\{\mathbf{k}s\}$, such that the trial wave function for the ground state can be written in the diagonal form of $\{\mathbf{k}s\}$ as
\begin{equation}\label{Eq:newwavef}
\begin{aligned}
|\Psi \rangle =\left(a_0+\sum _{\mathbf{k}s}a_{\mathbf{k}s}d_{\mathbf{k}s}^{\dagger}\gamma_{\mathbf{k}s}\right)|\Psi_0\rangle.
\end{aligned} 
\end{equation}
$a_{0}$, $a_{\mathbf{k}s}$ are all numbers and they are the variational parameters to be determined through self-consistent calculations. 

The energy of total Hamiltonian in the variational state $|\Psi \rangle $ shall be
\begin{equation}\label{Eq:energy}
\begin{aligned}
E=\frac{\langle\Psi |H|\Psi\rangle}{\langle\Psi|\Psi\rangle},
\end{aligned}
\end{equation}
where $\langle\Psi|\Psi\rangle=a_0^2+\sum_{\mathbf{k}s}a_{\mathbf{k}s}^2$=1 according to the wavefunction normalization condition. 



\begin{figure}[t]
	\begin{center}
		\includegraphics[scale=0.4, bb=280 60 400 517]{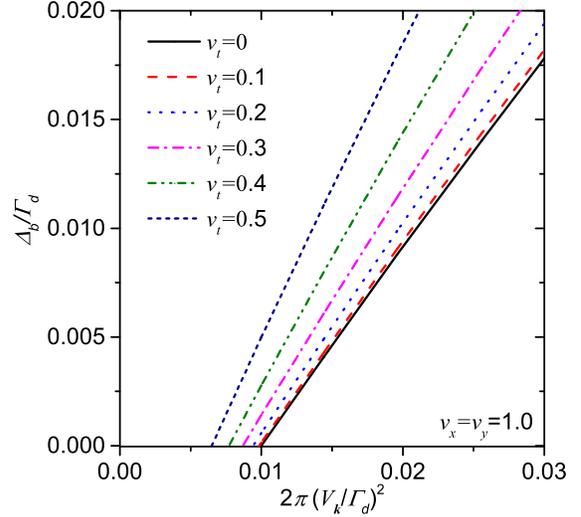}
	\end{center}
	\caption{(Color online). The results of binding energy for $v_x=v_y=1.0$ at $\mu=0$ with various values of $v_t$. The impurity energy level is chosen as $\epsilon_d = -0.01\Gamma_d$. When $v_t=0$, the magnetic impurity and the conduction electrons form bound states only if $2\pi(V_\mathbf{k}/\Gamma_d)^2 > \frac{|\epsilon_d|}{\Gamma_d}$\cite{Feng2010}. Thus the critical value of hybridization shall be $V_c = \sqrt{|\epsilon_d|\Gamma_d/(2\pi)}$.  As $v_t$ increases, there still exists a critical value of hybridization $V_c$, and it decreases as the Dirac cone is more and more strongly tilted.    } \label{Fig:5_binding}
\end{figure}

Then the total energy of the tilted Dirac system with a magnetic impurity in the trial state $|\Psi \rangle$ writes

\begin{equation}\label{Eq:totalE}
\begin{aligned}
E&=\frac{\sum_{\mathbf{k}s} \left[ (E_0 - \epsilon_{\mathbf{k}s} +\mu)a_{\mathbf{k}s}^2 + 2V_\mathbf{k}a_0a_{\mathbf{k}s} +(\epsilon_{\mathbf{k}s}-\mu)a_0^2 \right]}
{a_0^2+\sum_{\mathbf{k}s}a_{\mathbf{k}s}^2}.
\end{aligned}
\end{equation}

The variational principle requires that $\partial E/\partial a_0=\partial E/\partial a_{\mathbf{k}}=0$, which will lead us to two equations below:
\begin{equation}\label{Eq:threeab}
\begin{aligned}
& (E-\sum_{\mathbf{k}s}(\epsilon_{\mathbf{k}s}-\mu))a_0 = \sum_{\mathbf{k}s}V_\mathbf{k} a_{\mathbf{k}s},\\
& (E-E_0+(\epsilon_{\mathbf{k}s}-\mu))a_{\mathbf{k}s} = V_\mathbf{k} a_0.\\
\end{aligned}
\end{equation}

We then obtain the self-consistent equation
\begin{equation}\label{Eq:selfConsis}
\begin{aligned}
 \epsilon_d - \mu - \Delta_b = \sum_{\mathbf{k}s} \frac{V_\mathbf{k}^2}{\epsilon_{\mathbf{k}s} -\mu- \Delta_b },  \\
\end{aligned}
\end{equation}
$\Delta_b=E_0-E$ is the binding energy. If $\Delta_b>0$, the hybridized state has lower energy and is more stable than the bare state.
$\Delta_b$ can be obtained by numerically solving Eq. \ref{Eq:selfConsis}, and $a_0$ and $a_\mathbf{k}$ can be calculated according to the relations
\begin{equation}\label{Eq:a0ak}
\begin{aligned}
a_0^2 + \sum_{\mathbf{k}s}a_{\mathbf{k}s}^2=1,\\
a_{\mathbf{k}s} = \frac{V_\mathbf{k}}{\epsilon_{\mathbf{k}s}-\mu-\Delta_b}a_0.
\end{aligned}
\end{equation}

If $v_x= v_y= 1.0$ and $v_t =0$ the Dirac cone is not tilted at all, the band structure given in Eq. \ref{Eq:tilt_dispersion} is isotropic in the momentum space and the binding energy shall be exactly the same as that in a 2D helical metal\cite{Feng2010}. If $\mu=0$, the DOS is zero, such that the hybridization has a critical value $V_c$, below which the system has no positive binding energy. The results of the binding energy for $v_x=v_y=1.0$ with various $v_t$ values are given in Fig. \ref{Fig:5_binding}. 
The impurity energy level is chosen as $\epsilon_d = -0.01\Gamma_d$. When $v_t=0$, the magnetic impurity and the conduction electrons form bound states only if $2\pi(V_\mathbf{k}/\Gamma_d)^2 > \frac{|\epsilon_d|}{\Gamma_d}$\cite{Feng2010}. Thus the critical value of hybridization shall be $V_c = \sqrt{|\epsilon_d|\Gamma_d/(2\pi)}$.  As $v_t$ increases, there still exist a critical value of hybridization $V_c$, since the DOS at the Fermi energy still vanishes for $v_t<v_y$. However, $V_c$ decreases as the Dirac cone is more strongly tilted, indicating that the tilted Dirac system forms a bound state more easily than the Dirac cones which are not tilted.  
For a more complicated case when $v_x \neq v_y$, if the DOS at $\mu=0$ is still zero, there should exist a critical value of hybridization $V_c$, since the existence of a critical value merely depends on the DOS at the Fermi energy. The values of $V_c$ is determined by the velocities $v_i$ ($i=x,y,t$).

When $\mu\neq 0$, the DOS at the Fermi energy becomes finite, so there exists positive binding energy for arbitrary $V_k$ values. While $v_t> 0$, the band structure of the Dirac cone is tilted along the $k_y$ axis, and if $v_t$ is larger than $v_y$, the Dirac cone is so strongly tilted that the DOS at the Fermi energy for $\mu=0$ becomes finite. In this case the magnetic impurity and the conduction electrons always form a bound state.

\section{Spin-spin correlation}\label{Sec:sscorr}

In this section, we study about the spin-spin correlation between the magnetic impurity and the conduction electrons.
The spin operators of magnetic impurity and conduction electrons are defined as $\mathbf{S_d}=\frac{1}{2} d^{\dagger}\sigma d$, $\mathbf{S_c}=\frac{1}{2} c^{\dagger}\sigma c$, where $\sigma$ is the spin-Pauli Matrix.
The Fourier transformations of the conduction electrons read
$c_{\sigma }(\bf{r})=\frac{1}{\sqrt{N}}\sum_{\bf{q}} e^{\text{i\bf{qr}}}c_{\text{\bf{q}$\sigma$}} ;\text{    }c_{\sigma }{}^{\dagger }(r)=\frac{1}{\sqrt{N}}\sum
_{\bf{q}} e^{-\text{i\bf{qr}}}c_{\text{\bf{q}$\sigma$}}{}^{\dagger }$.
We choose the position of magnetic impurity as $\mathbf{r}=0$, and consider about spin-spin correlation $\mathbf{J}_{uv}(\mathbf{r})= \langle S_{c}^u (\mathbf{r})S_d^v(0)\rangle$ on the $x-y$ plane, where $\mathbf{r}$ is the location of the conduction electron. Here $u,v=x,y,z$ and $\langle \cdots \rangle$ denotes the ground state average.

\begin{figure}[t]
	\begin{center}
		\includegraphics[scale=0.53, bb=180 100 400 775]{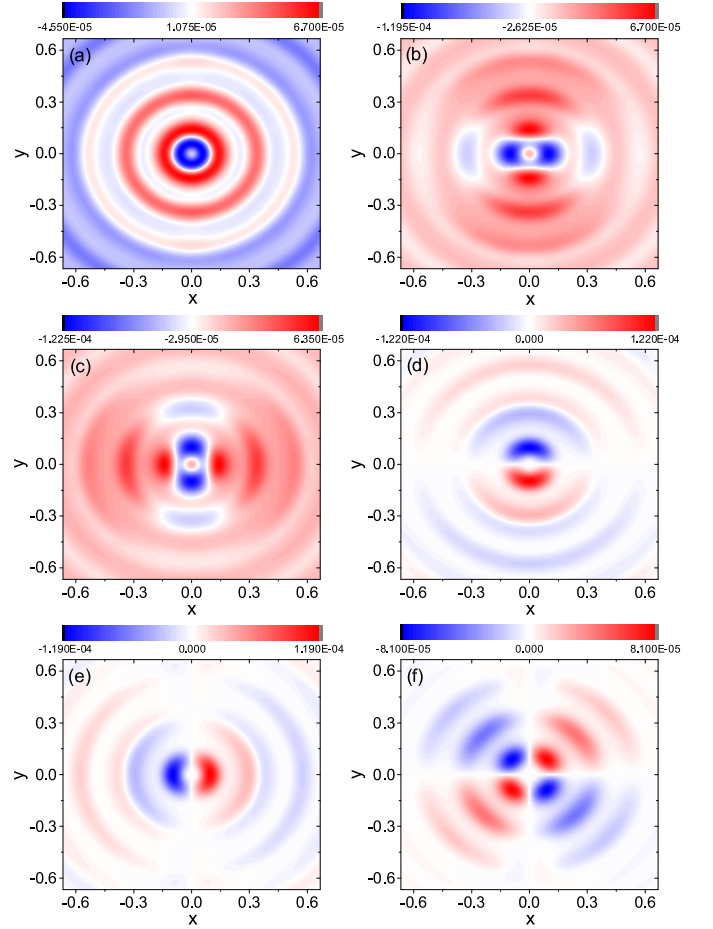}
	\end{center}
	\caption{(Color online). The results of $J_{uv}(\mathbf{r})\times r^2$ for $v_x = v_y = 1.0$ and $v_t=0$. (a) $r^2 J_{zz}(\mathbf{r})$, (b) $r^2 J_{xx}(\mathbf{r})$, (c) $r^2 J_{yy}(\mathbf{r})$, (d) $r^2 J_{xz}(\mathbf{r})$, (e) $r^2 J_{yz}(\mathbf{r})$, (f) $r^2 J_{xy}(\mathbf{r})$.  } \label{Fig:1_110}
\end{figure}

The spin-spin correlation function is evaluated for relatively small $v_t$ and for $\mu \neq 0$. In this case, the DOS at half-filling still vanishes, but the DOS is significant when $\mu\neq 0$ such that the binding energy $\Delta_b$ is always positive. This means that the magnetic impurity and the conduction electrons always form a bound state.
The diagonal terms and the nonzero off-diagonal terms of the spin-spin correlation in coordinate space are given by
\begin{equation}\label{Eq:sscorr}
\begin{aligned}
\mathbf{J}_{zz}(\mathbf{r})&= -\frac{1}{8}  \left| \mathcal{A(\mathbf{r})} \right|^2 + \frac{1}{16} \left| \mathcal{B(\mathbf{r})} \right|^2  +\frac{1}{16} \left| \mathcal{C(\mathbf{r})} \right|^2  ,\\
\mathbf{J}_{xx}(\mathbf{r})&=  -\frac{1}{8}  \left| \mathcal{A(\mathbf{r})} \right|^2 - \frac{1}{8} \text{Re}\left[ \mathcal{B^*(\mathbf{r})} \mathcal{C(\mathbf{r})} \right] \\
\mathbf{J}_{yy}(\mathbf{r})&= -\frac{1}{8}  \left| \mathcal{A(\mathbf{r})} \right|^2 + \frac{1}{8} \text{Re}\left[ \mathcal{B^*(\mathbf{r})} \mathcal{C(\mathbf{r})} \right] \\
\mathbf{J}_{xz}(\mathbf{r})&= -\frac{1}{8}\text{Re}\left[ \mathcal{A^*(\mathbf{r})} \mathcal{B(\mathbf{r})}  \right]+
                              \frac{1}{8}\text{Re}\left[ \mathcal{A^*(\mathbf{r})} \mathcal{C(\mathbf{r})}  \right], \\
\mathbf{J}_{yz}(\mathbf{r})&=  \frac{1}{8}\text{Im}\left[ \mathcal{A^*(\mathbf{r})} \mathcal{B(\mathbf{r})}  \right]+
                              \frac{1}{8}\text{Im}\left[ \mathcal{A^*(\mathbf{r})} \mathcal{C(\mathbf{r})}  \right],  \\
\mathbf{J}_{xy}(\mathbf{r})&= \frac{1}{8}\text{Im}\left[ \mathcal{B^*(\mathbf{r})} \mathcal{C(\mathbf{r})}  \right], \\
\end{aligned}
\end{equation}
where $\mathcal{A}(\mathbf{r})= \sum_{\mathbf{k}s} e^{i\mathbf{k\cdot r}}a_{\mathbf{k}s}$,  $\mathcal{B}(\mathbf{r})= \sum_{\mathbf{k}s} \text{sgn}(s) e^{i(\mathbf{k\cdot r}+\theta_{\mathbf{k}})}a_{\mathbf{k}s}$, $\mathcal{C}(\mathbf{r})= \sum_{\mathbf{k}s} \text{sgn}(s) e^{i(\mathbf{k\cdot r}-\theta_{\mathbf{k}})}a_{\mathbf{k}s}$.

\begin{figure}[t]
	\begin{center}
		\includegraphics[scale=0.53, bb=180 100 400 775]{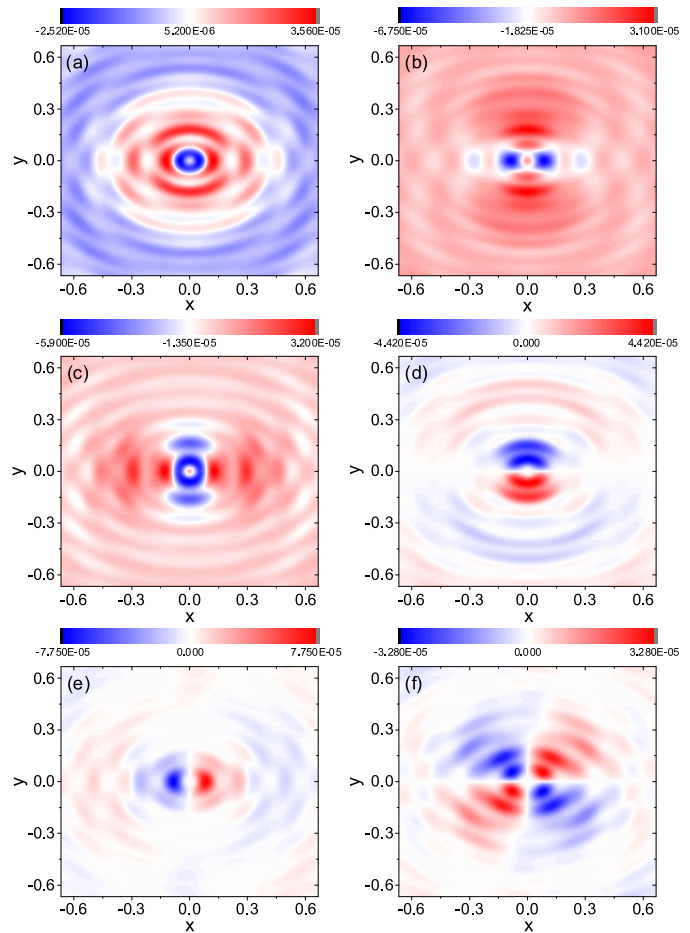}
	\end{center}
	\caption{(Color online). The results of $J_{uv}(\mathbf{r})\times r^2$ for $v_x = v_y = 1.0$ and $v_t=0.5$. (a) $r^2 J_{zz}(\mathbf{r})$, (b) $r^2 J_{xx}(\mathbf{r})$, (c) $r^2 J_{yy}(\mathbf{r})$, (d) $r^2 J_{xz}(\mathbf{r})$, (e) $r^2 J_{yz}(\mathbf{r})$, (f) $r^2 J_{xy}(\mathbf{r})$.  } \label{Fig:2_1105.eps}
\end{figure}

\begin{figure}[t]
	\begin{center}
		\includegraphics[scale=0.53, bb=180 100 400 775]{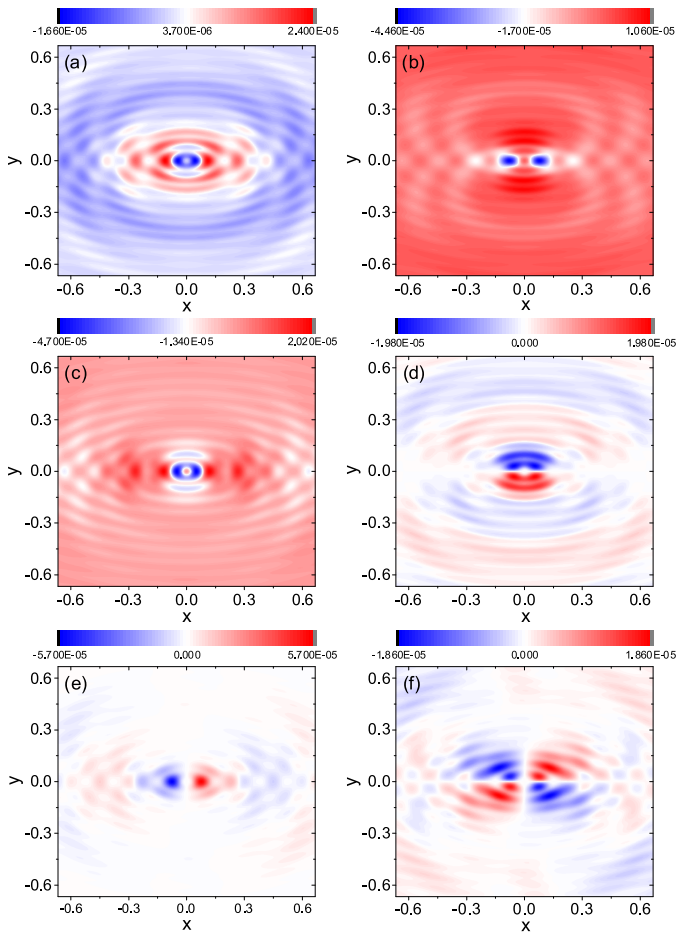}
	\end{center}
	\caption{(Color online). The results of $J_{uv}(\mathbf{r})\times r^2$ for $v_x =1.0$, $ v_y = 0.8$ and $v_t=0.5$. (a) $r^2 J_{zz}(\mathbf{r})$, (b) $r^2 J_{xx}(\mathbf{r})$, (c) $r^2 J_{yy}(\mathbf{r})$, (d) $r^2 J_{xz}(\mathbf{r})$, (e) $r^2 J_{yz}(\mathbf{r})$, (f) $r^2 J_{xy}(\mathbf{r})$.  } \label{Fig:4_10805}
\end{figure}

In Fig. \ref{Fig:1_110} - Fig. \ref{Fig:4_10805} we show the patterns of spatial spin-spin correlation between the magnetic impurity and conduction electrons in the $x-y$ plane, for different values of $v_x$, $v_y$ and $v_t$. For all the cases, the Dirac cone is weakly tilted that the DOS at the Dirac point is still zero. The various components of spin-spin correlation show spatial oscillations and decay with respect to the displacement $\mathbf{r}$. In order to investigate the patterns more clearly, we show $J_{uv}(\mathbf{r})\times r^2$ instead of $J_{uv}(\mathbf{r})$, and the length unit is $1/\Gamma_d$ where $\Gamma_d$ is the energy cut-off.
The spin-spin correlation between the magnetic impurity and a conduction electron of distance $\mathbf{r}$ follows a power law decay $1/r^{d}$ if $r<\xi_K$, and $1/r^{d+1}$ if $r>\xi_K$, with $\xi_K$ the Kondo coherence length and $d$ the dimensionality of the host material\cite{ishii1978,Barzykin1998,Borda2007}.
In fact the binding energy $\Delta_b$ shall take different values while $v_x$, $v_y$ or $v_t$ changes. Here for simplicity, we may fix $\Delta_b$ as a constant value since the change of the spatial spin-spin correlation is our major concern. The parameter we use in this section is $V_k = 0.05 \Gamma_d$,  $\Delta_b=0.1\Gamma_d$, $\mu=-0.1\Gamma_d$.
We may find through simple calculation that the off-diagonal terms of $J_{uv}(\mathbf{r})$ have the relation that $J_{xz}(\mathbf{r})=-J_{zx}(\mathbf{r})$, $J_{yz}(\mathbf{r})=-J_{zy}(\mathbf{r})$ and $J_{xy}(\mathbf{r})=J_{yx}(\mathbf{r})$, so only $J_{xz}(\mathbf{r})$, $J_{yz}(\mathbf{r})$ and $J_{xy}(\mathbf{r})$ are explicitly given in Fig. \ref{Fig:1_110} - Fig. \ref{Fig:4_10805}.

Shown in Fig. \ref{Fig:1_110} are the results of $J_{uv}(\mathbf{r})$$(u,v=x,y,z)$ while $v_x = v_y = 1.0$ and $v_t=0$. The spatial patterns of all the six components of the spin-spin correlation shall be exactly the same as those given in a 2D helical metal\cite{Feng2010}.
$J_{zz}(\mathbf{r})$ shown in Fig. \ref{Fig:1_110} (a) is antiferromagnetic around the magnetic impurity, and is isotropic in the coordinate space. Both $J_{xx}(\mathbf{r})$ in Fig. \ref{Fig:1_110} (b) and $J_{yy}(\mathbf{r})$ in Fig. \ref{Fig:1_110} (c) are also dominated by antiferromagnetic correlation around the impurity location, but is spatially anisotropic along the $x$- or $y$-axis. $J_{xz}(\mathbf{r})$ plotted in Fig. \ref{Fig:1_110} (d) shows more interesting behavior. Around the magnetic impurity, the correlation is antiferromagnetic while $y>0$ and ferromagnetic while $y<0$, and is zero along the $y$-axis. $J_{yz}(\mathbf{r})$ in Fig. \ref{Fig:1_110} (e) shows the same behavior as $J_{xz}(\mathbf{r})$ if we exchange the real space coordinate $x\rightarrow y$ and $y\rightarrow x$. $J_{xy}(\mathbf{r})$ is plotted in Fig. \ref{Fig:1_110} (f). It is ferromagnetic while $xy>0$ and antiferromagnetic when $xy<0$, and is zero along both the $x$- and $y$- axes. While $v_x=v_y$ and $v_t=0$, the dispersion relation of the Dirac cone is isotropic in the momentum space, and hence the various components of spin-spin correlation between the magnetic impurity and conduction electrons show highly symmetric pattern. However, when $v_t$ term becomes finite, the Dirac cone is tilted along the $y$-axis, and accordingly the $J_{uv}(\mathbf{r})$ ($u,v=x,y,z$) becomes highly anisotropic in the $x-y$ plane.

Shown in Fig. \ref{Fig:2_1105.eps} are the results of $J_{uv}(\mathbf{r})\times r^2$ for $v_x = v_y = 1.0$ and $v_t=0.5$. The band structure of the tilted Dirac cone is given in Fig. \ref{Fig:0_dispersion.pdf}, that the symmetry between the $k_x$- and $k_y$-axis are broken by the non-zero $v_t$ term. The broken symmetry in the momentum space also affect the patterns of spin-spin correlation in the real space, and the results are shown in Fig. \ref{Fig:2_1105.eps}. We can see that all the components of spin-spin correlation oscillates faster along the $y$-axis, and slower along the $x$-axis in comparison to those give in Fig. \ref{Fig:1_110}. The spatial spin-spin correlation shows clear interference patterns with large $r$.
$J_{zz}(\mathbf{r})$ shown in Fig. \ref{Fig:2_1105.eps} (a) becomes strongly anisotropic in real space.
Around the magnetic impurity, $J_{zz}(\mathbf{r})$ is still antiferromagnetic, but the correlation along the $x$- and $y$-axis oscillates in different periods. $J_{xx}(\mathbf{r})$ and $J_{yy}(\mathbf{r})$ are both squeezed along the $y$-axis, and the interference pattern emerges for large $r$. $J_{xx}(\mathbf{r})$ and $J_{yy}(\mathbf{r})$ given in Fig. \ref{Fig:2_1105.eps} (b) and (c) also show interference pattern when $r$ is away from the magnetic impurity location. For both of the spin-spin correlation components, the antiferromagnetic behavior around the magnetic impurity remains unchanged, but the oscillation on the $x$-, $y$-axis becomes slightly different. In Fig. \ref{Fig:2_1105.eps} (d) and (e), we show $J_{xz}(\mathbf{r})$
and $J_{yz}(\mathbf{r})$ which show much different patterns in comparison with those given in Fig. \ref{Fig:1_110} (d) and (e).
$J_{xz}(\mathbf{r})$ and $J_{yz}(\mathbf{r})$ are both squeezed along the $y$-axis, and show clear interference patterns near the $x$-axis while $r$ is large. $J_{xy}(\mathbf{r})$ given in Fig. \ref{Fig:2_1105.eps} (e) is the most interesting one. Besides the interference patterns for large $r$, it also shows different symmetry. When $v_t=0$ as shown in Fig. \ref{Fig:1_110} (f), the $J_{xy}(\mathbf{r})$ is always zero along the $x$- or $y$-axis, and the absolute value has a 4-fold rotational symmetry. However when $v_t\neq 0$, the $J_{xy}(\mathbf{r})$ is still zero along the $x$-axis, but becomes non-zero along the $y$-axis. The 4-fold rotational symmetry of the absolute value is also broken due to the tilting term.

In Fig. \ref{Fig:4_10805}, we shows the spin-spin correlation components while $v_x=1.0$, $v_y=0.8$ and $v_t=0.5$. Actually in the $8-pmmn$ borophene\cite{Lopez2016}, the typical value of the parameters are $v_x=0.89$, $v_y=0.67$ and $v_t=0.32$. Hence our choice of the $v_i$($i=x,y,t$) values will show spin-spin correlation patterns very close to those in a $8-pmmn$ borophene. Besides the $v_t$ term which tilts the Dirac cone along the $k_y$ axis, the velocity along the $k_x$- and $k_y$-axis becomes distinct, and this will add extra anisotropy in the momentum space.
In general, we can easily find that the components of the spatial spin-spin correlation are more strongly squeezed than those in Fig. \ref{Fig:2_1105.eps}. Here we set $v_x > v_y$ that the anisotropy along the $x$- and $y$-axis is enhanced by the velocity terms. We can see that the spin-spin correlation decays and oscillates much faster along the $y$-axis and slower along the $x$-axis.
$J_{zz}(\mathbf{r})$ shown in Fig. \ref{Fig:4_10805} (a) becomes more strongly anisotropic in the coordinate space.
Around the magnetic impurity, $J_{zz}(\mathbf{r})$ is still antiferromagnetic, but the correlation along the $x$-axis oscillates much slower than that along the $y$-axis. This is caused by the distinct velocity along the $k_x$, $k_y$-axis. In contrast, if we choose $v_x<v_y$, the velocity difference will compensate the anisotropy caused by the tilting term, and shows spin-spin correlation patterns more close to those given in Fig. \ref{Fig:1_110}.  $J_{xx}(\mathbf{r})$ and $J_{yy}(\mathbf{r})$ given in Fig. \ref{Fig:4_10805} (b) and (c) are more strongly squeezed along the $y$-axis, and show more clear interference pattern when $r$ is away from the magnetic impurity location. The antiferromagnetic nature remains unchanged, but the oscillation along the $x$-, $y$-axis show completely distinct patterns. When $v_x=v_y$ and $v_t=0$ as given in Fig. \ref{Fig:1_110}, $J_{xx}(\mathbf{r})$ and $J_{yy}(\mathbf{r})$ if we rotate the coordinate space by $90^\circ$. However, this symmetry is completely broken by the tilting term and the distinct velocities along the $x$-, $y$-axis.
In Fig. \ref{Fig:4_10805} (d) and (e), we show $J_{xz}(\mathbf{r})$
and $J_{yz}(\mathbf{r})$ which are more strongly squeezed along the $y$-axis, and shows clear interference patterns near the $x$-axis for large $r$. $J_{xy}(\mathbf{r})$ is given in Fig. \ref{Fig:4_10805} (e). We can see that the 4-fold rotational symmetry of the absolute value is completely destroyed. The $x-y$ spin-spin correlation is still zero along the $x$-axis, but is clearly non-zero along the $y$-axis.

\section{conclusions}\label{conclusion}
In this paper we utilize the variational method study the Kondo screening of a spin-1/2 magnetic impurity in tilted Dirac surface states at the large-$U$ limit. The host material is described by a tilted Dirac cone in two dimensions. The Kondo screening in topological semimetals using the same trial wavefunction method had been studied in Ref. \cite{Jinhua2015}. In order to see the spatial changes of spin-spin correlation, we choose two sets of $v_i$($i=x,y,t$) parameters, they are: (1) $v_x=v_y=1.0$, $v_t=0.5$, (2) $v_x = 1.0$, $v_y=0.8$, $v_t=0.5$ and compare the results with the counterparts in a 2D helical metal while $v_x=v_y=1.0$, $v_t=0$\cite{Feng2010}.
When the Dirac cone is slightly tilted ($v_t<v_x$, $v_y$), the DOS at a charge neutral point still vanishes as in graphene, so there exist a critical value of hybridization $V_c$. The magnetic impurity and conduction electrons form a bound state only if $V_\mathbf{k} > V_c$. If the Fermi surface is tuned away from the Dirac point, then the magnetic impurity and conduction electrons will always form a bound state for arbitrary $V_\mathbf{k}$.
If a finite $v_t$ term is added, the Dirac cone is tilted along the $k_y$-axis. The components of the spatial spin-spin correlation oscillates with different period along the $x$- or $y$-axis, and show more anisotropic patterns. The tilting of the Dirac cone does not change the signs of correlation close to the magnetic impurity, but interference patterns show up while $r$ is large.

So far, we have only studied the effect of a single magnetic impurity in tilted Dirac fermion systems with spin-orbit coupling in two dimensions. The 3D tilted Dirac/Weyl fermion systems should exhibit similar behaviors to those of the 2D tilted Dirac systems. However, the spin-spin correlation is expected to show more rich patterns due to an extra dimension and these will be investigated in our future work.

\section{Acknowledgement}
This work is supported in part by NSFC (under Grant No. 11604166) and the K.C. Wong Magna Fund in Ningbo University.
L. Li is supported by the NSFC (under Grant No. 11604138),
D.-H. Xu is supported by the NSFC (under Grant No. 11704106) and the Chutian Scholars Program
in Hubei Province.

\bibliographystyle{apsrev4-1}
\bibliography{ref}

\end{document}